\documentclass[conference]{IEEEtran}
\IEEEoverridecommandlockouts
\usepackage{cite}
\usepackage[numbers]{natbib}
\usepackage{amsmath,amssymb,amsfonts}
\usepackage{graphicx}
\usepackage{textcomp}
\usepackage{xcolor}
\usepackage{booktabs}
\usepackage{tabularx}
\usepackage{multirow}
\usepackage[table, xcdraw]{xcolor}
\usepackage{algorithm}
\usepackage{algpseudocode}
\usepackage[colorlinks=true, linkcolor=blue, citecolor=blue, urlcolor=blue]{hyperref}

\usepackage{amsmath, amssymb}

\def\BibTeX{{\rm B\kern-.05em{\sc i\kern-.025em b}\kern-.08em
    T\kern-.1667em\lower.7ex\hbox{E}\kern-.125emX}}

\makeatletter
\newcommand*\titleheader[1]{\gdef\@titleheader{#1}}
\AtBeginDocument{%
  \let\st@red@title\@title
  \def\@title{%
    \bgroup\normalfont\normalsize\centering\@titleheader\par\egroup
    \vskip1ex\st@red@title}
}
\makeatother
\newcommand{\sys}{BiTMedViT}

\title{Invited Paper: BitMedViT: \\Ternary-Quantized Vision Transformer for\\ Medical AI Assistants on the Edge}

\titleheader{\vspace{-25pt}Accepted at 2025 IEEE/ACM International Conf. on Computer-Aided Design (ICCAD) Oct. 26-30 2025, Munich, DE}

\author{\IEEEauthorblockN{
Mikolaj Walczak,
Uttej Kallakuri,
Edward Humes,
Xiaomin Lin,
Tinoosh Mohsenin}
\IEEEauthorblockA{\textit{Department of Electrical and Computer Engineering} \\
\textit{Johns Hopkins University}\\
\{mwalcza1, ukallak1, ehumes2, xlin52, tinoosh\}@jh.edu}}

\begin{document}

\vspace{0.5cm}

\maketitle

\begin{abstract}

Vision Transformers (ViTs) have demonstrated strong capabilities in interpreting complex medical imaging data.
However, their significant computational and memory demands pose challenges for deployment in real-time, resource-constrained mobile and wearable devices used in clinical environments. We introduce, \sys{},  a new class of Edge ViTs serving as medical AI assistants that perform structured analysis of medical images directly on the edge. \sys{} utilizes ternary-quantized linear layers tailored for medical imaging and combines a training procedure with multi-query attention, preserving stability under ternary weights with low-precision activations. Furthermore, \sys{} employs task-aware distillation from a high-capacity teacher to recover accuracy lost due to extreme quantization. Lastly, we also present a pipeline that maps the ternarized ViTs to a custom CUDA kernel for efficient memory bandwidth utilization and latency reduction on the Jetson Orin Nano. 
Finally, \sys{} achieves 86\% diagnostic accuracy (89\% SOTA) on MedMNIST across 12 datasets, while reducing model size by 43$\times$, memory traffic by 39$\times$, and enabling 16.8~ms inference at an energy efficiency up to 41$\times$ that of SOTA models  at 183.62~GOPs/J on the Orin Nano.
Our results demonstrate a practical and scientifically grounded route for extreme-precision medical imaging ViTs deployable on the edge, narrowing the gap between algorithmic advances and deployable clinical tools. Github is available at \href{https://github.com/M-iki/BitMedViT}{https://github.com/M-iki/BitMedViT}
 
\end{abstract}
\vspace{-2ex}

\section{Introduction}\label{sec:intro}
\vspace{-1ex}
Machine learning for medical imaging \cite{chen2025review, kim2022transfer, zhang2024segment, rayed2024deep} and disease detection \cite{guan2024federated} are rapidly advancing fields with the potential to transform healthcare by enabling real-time, automated analysis of imaging modalities such as X-rays, MRIs and CT scans. This automation supports accurate disease identification, improves diagnostic precision and accelerates clinical decision-making. Historically, Convolutional Neural Networks (CNNs), especially those with residual architectures like ResNet\cite{he2016deep}, have demonstrated strong performance. 
Recent breakthroughs \cite{manzari2023medvit, yue2024medmamba, manzari2025medical} have advanced medical image classification by surpassing traditional CNN accuracy through sophisticated architectures such as Vision Transformers (ViTs) \cite{dosovitskiy2020image} and Vision Mamba~\cite{zhu2024vision,aalishah2025medmambalite}. 

\begin{figure}
    \centering
    \includegraphics[width=\linewidth]{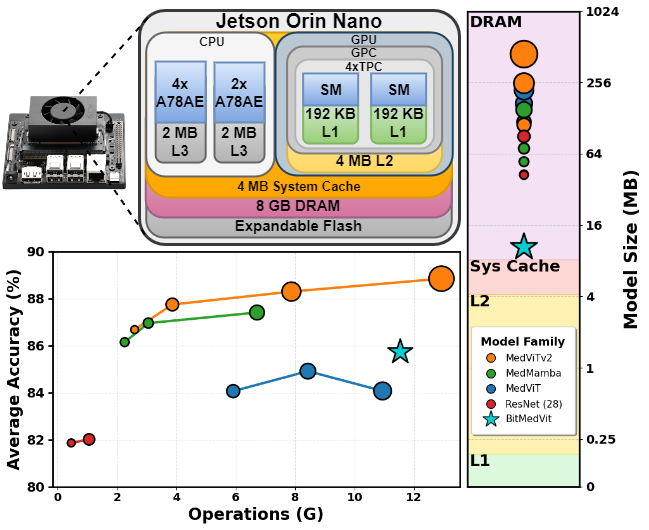}
    \vspace{-5ex}
    \caption{Jetson Orin Nano hierarchy and SOTA medical image classification model comparisons for average accuracy on the MedMNIST dataset vs operations and model size compared to the memory available within the GPU consisting of 4 TPCs each with two SMs containing their own L1 cache. Marker size corresponds to the parameter count within each model.}
    \label{fig:1}
    \vspace{-4ex}
\end{figure}

Despite these advances, deploying ViTs in clinical environments remains challenging due to their high computational and memory demands, which restrict adoption in clinical settings with limited hardware resources or unreliable connectivity. Furthermore, transmitting sensitive medical data over wireless networks introduces privacy and security risks~\cite{al2012security} - a major concern given HIPAA and other stringent medical data regulations~\cite{marks2023ai}. To address these issues, we investigate deploying these models on the resource-constrained NVIDIA Jetson Orin Nano edge device. Equipped with Arm Cortex-A78AE CPU and an NVIDIA Ampere GPU, the Orin Nano is tailored for low-power, resource-constrained environments. Deploying models optimized for its core computing architecture ensures efficient resource utilization, reduced power consumption, and optimal performance within strict energy budgets. As shown in Figure \ref{fig:1}, the state-of-the-art medical classification models, even under ideal settings, exceed the low memory hierarchy (L1, L2, L3) limits, causing high latency, low throughput, and increased energy use due to the reliance on external accesses to high-latency DRAM. To overcome these deployment constraints, recent research has emphasized model compression techniques such as pruning \cite{kallakuri2024resource, kallakurik2025enabling, hosseini2021cyclic, uttejatecs2025, manjunath2021energy}, quantization \cite{mazumder2023reg}, and Knowledge Distillation (KD) \cite{aalishah2025mambalitesr, rashid2024tinyvqa, rashid2025hac} that reduce model size and inference costs while maintaining accuracy. KD has proven especially effective, and when combined with quantization, it yields lightweight models retaining high performance at a fraction of the original footprint, which is ideal for edge medical AI applications. In ViTs, further reductions in parameters and memory can be achieved by replacing standard Multi-Head Self-Attention with Multi-Query Attention, significantly cutting computational overhead with minimal impact on accuracy \cite{kitaev2020reformer, touvron2023llama, anil2023palm}.

\begin{figure*}[ht]
    \centering
    \includegraphics[width=0.91\linewidth]{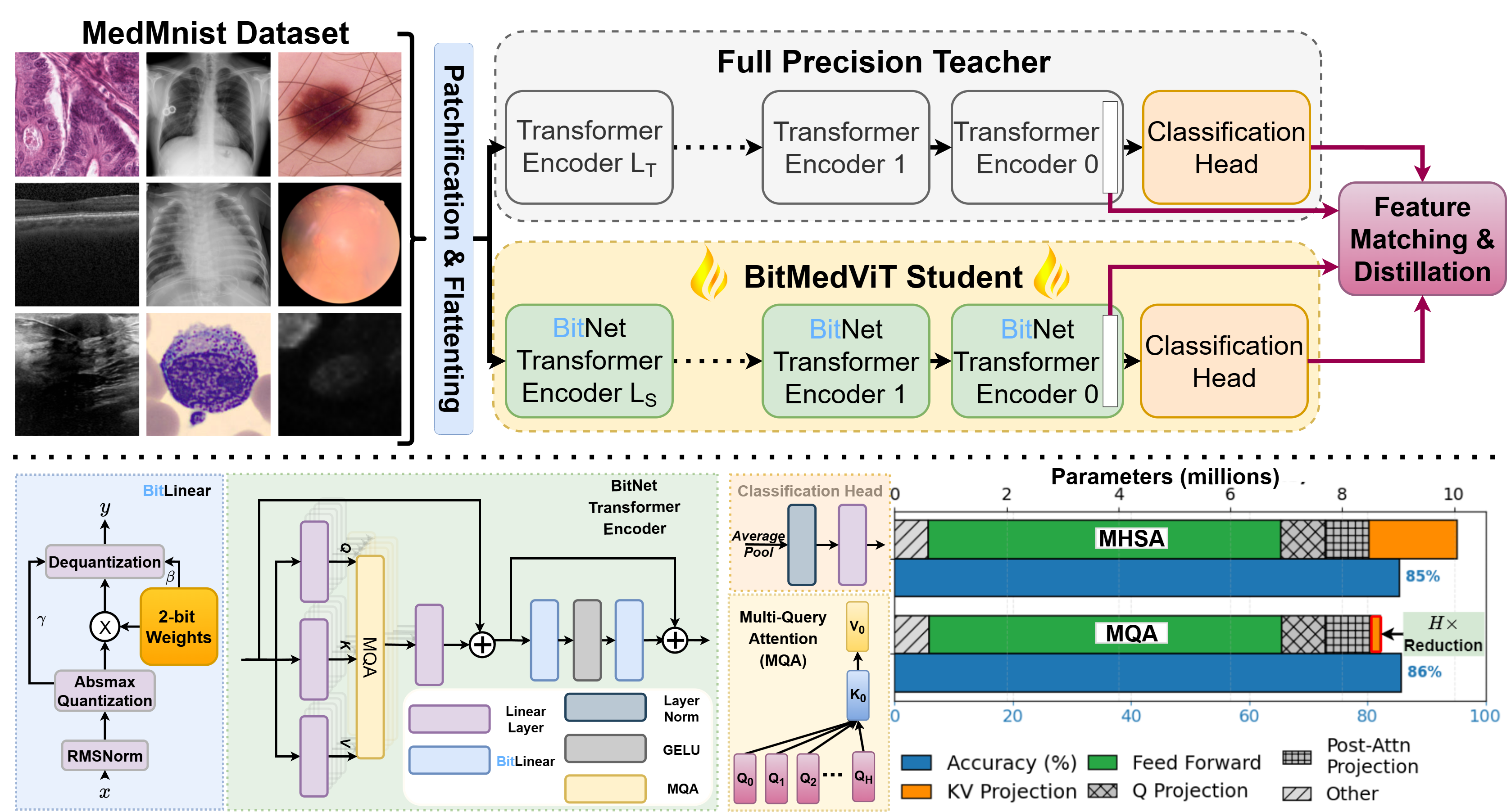}
    \vspace{-2ex}
    \caption{Model training paradigm and \sys{} Architecture. Experiments illustrate the overall parameter breakdown and reduction when comparing Multi-Head Self-Attention (MHSA) and Multi-Query Attention (MQA) mechanisms within \sys{}. The blue plots depict the average validation accuracy achieved when training 12 student models  across each of the MedMNIST2D~\cite{medmnistv1, medmnistv2} datasets rounded to the nearest percentage point after 100 epochs.}
    \label{fig:2}
    \vspace{-4ex}
\end{figure*}

In this work, we introduce \sys{}, which integrates KD with extreme quantization to enhance model compression. We focus on 2-bit ternary quantization of feed-forward layers using the BitNet-1.58B framework \cite{ma2024era} and are one of the first works to explore MQA in ViTs to compress the model footprint for medical image classification. We evaluate our method for accuracy and efficiency on edge hardware, benchmarking against MedViTV2~\cite{manzari2025medical} deployed on the NVIDIA Jetson Orin Nano with custom CUDA kernels optimized for performance.

Our key contributions are summarized as follows:
\begin{itemize}
\item We evaluate the efficacy of MQA in ViTs for medical image classification, demonstrating more efficient parameter use while preserving accuracy.
\item We apply feature and logit distillation from a high-performing state-of-the-art teacher model  and propose a distilled student \sys{} for medical applications that enables robust classification .
\item We integrate BitNet linear layers within \sys{}, enabling the training of scalable, memory- and parameter-efficient edge-based ViTs. 
\item We develop a optimized CUDA kernel compatible with TensorRT and compatible with Nvidia Ampere GPU architectures, found within the Jetson Orin Nano, enabling real-time inference and efficient memory utilization for edge-based medical AI applications.
\end{itemize}
\section{Related Works}

\subsection{Knowledge Distillation for Medical ViTs}

Knowledge distillation (KD) is a proven strategy for compressing large models, particularly effective in data-scarce medical imaging scenarios. Feature-based KD enhances student learning by aligning internal representations with those of a powerful teacher model. Medical-specific distillation methods such as \cite{sevinc2025distillation, park2022self} address limited and imbalanced data, while quantization-aware techniques \cite{kim2019qkd, zhu2023quantized} further reduce inference cost. \textbf{\textit{\sys{}}} builds on these by combining feature distillation with low-bit quantization to produce accurate, lightweight ViTs for edge devices.
\vspace{-1ex}

\subsection{Multi-Query Attention in Vision Models}
Multi Query Attention (MQA) reduces attention complexity by sharing key-value pairs across queries, offering significant efficiency gains over Multi-Head Self Attention (MHSA). Although adopted in Large Language Models (LLMs) \cite{touvron2023llama, anil2023palm}, MQA has seen little application in ViTs. Ainslie et al. \cite{ainslie2023gqa} show that MHSA architectures can be converted to MQA with minimal changes. \sys{} is among the first to evaluate MQA in ViTs for medical imaging.
\vspace{-1ex}

\subsection{Ternary ViTs and Edge Deployment}
Ternary quantization reduces weights and activations to three discrete values, balancing efficiency and accuracy. Prior works like Tervit \cite{xu2022tervit} and BitNet-ViT \cite{yuan2024vit} demonstrate that ViTs can be effectively quantized with minimal accuracy loss. However, these methods often overlook deployment feasibility on constrained hardware. \sys{} builds on the BitNet-1.58B framework \cite{ma2024era}, integrating BitLinear ternary layers and demonstrating practical, low-latency deployment on edge devices like the Jetson Orin Nano. Furthermore, deploying ViTs on edge devices requires tight alignment between model design and hardware constraints. While frameworks like BitNet-Efficient \cite{wang2025bitnet} and FPGA-based ternary transformers \cite{yin2025tereffic} demonstrate efficient execution, they target general-purpose or LLM scenarios. Prior works have also explored KD, quantization, and ternary ViTs for efficient model compression, but often lacks real-world deployment on medical setting. In addition, MQA is common in LLMs but is rarely applied to ViTs. Our proposed work bridges these gaps by combining feature distillation, extreme ternary quantization, and MQA into a compact ViT pipeline, optimized and deployed with custom CUDA kernels for real-time inference on edge devices like the Jetson Orin Nano.




\section{Proposed Approach}


\subsection{\sys{} Architecture}

The \sys{} architecture is based on the traditional ViT~\cite{dosovitskiy2020image}, parameterized by the number of attention heads~$H$, transformer layers~$L$, and patch embedding dimension~$E_d$. The post-attention Feed-Forward Networks (FFNs) employ an expansion factor of~4, yielding an FFN dimension of $E_{\text{ff}} = 4 \times E_d$. In this work, we fix the configuration to $L = 3$, $H = 8$, and $E_d = 512$, providing a balanced trade-off between model capacity for medical image analysis and a reduced overall parameter count. Prior to deployment on the Orin Nano GPU platform, it is essential to ensure the model is sufficiently compact to meet stringent memory constraints before hardware-specific optimizations are applied. As illustrated in Fig.~\ref{fig:2}, a parameter breakdown of the conventional MHSA architecture under this configuration shows that the FFN layers and key--value projections contribute the largest share of parameters. This motivates our focus on compressing these components to reduce overall memory footprint.


\subsubsection{Attention Layer Compression}

Transformer architectures rely heavily on the attention mechanism, which computes context-aware representations through learned input feature linear projections. This mechanism is formally expressed as
\vspace{-2ex}
\begin{equation}
\mathrm{Attention}(Q, K, V) = \mathrm{softmax}\left( \frac{QK^\top}{\sqrt{d_k}} \right) V,
\label{eq:self_attention}
\vspace{-1ex}
\end{equation}

where the queries (\(Q\)), keys (\(K\)), and values (\(V\)) are projections of the input (\(X\)). Moreover, the total embedding dimension in Equation\ref{eq:self_attention} is divided into multiple heads \(H\), each with separate \(Q\), \(K\), and \(V\) projection weights. While this increases parallelism and expressivity, the total number of parameters remains constant, as the projections are simply partitioned into smaller subspaces. MQA modifies this structure by sharing the key and value projections across all heads. Let $X\!\in\!\mathbb{R}^{N\times D}$ be the patch embeddings, with $H$ heads of size $d_h$ so $D=H d_h$.
The attention output for MHSA  can then be computed using Equation~\ref{eq:self_attention} with the projection weights of Q,K and V in a single head being
\vspace{-1ex}
\begin{equation}
W_h^Q,W_h^K,W_h^V\in\mathbb{R}^{D\times d_h}
\label{eq:mhsa_weights}
\vspace{-1ex}
\end{equation}

where $Q_h$, $K_h$ and $V_h$ are computed as $Q_h=XW_h^Q,\ K_h=XW_h^K,\ V_h=XW_h^V$. Within each head we further compute $Y_h=\mathrm{Attn}(Q_h,K_h,V_h)$, then $Y=\mathrm{Concat}_h(Y_h)W^O$.
The number of parameters in this case is, $3D d_h H$.

MQA on the other hand, maintains the per–head queries but shares the key and value projections across all the heads. Equation~\ref{eq:mhsa_weights} in this case can be rewritten as
\vspace{-1ex}
\begin{equation}
W_h^Q\in\mathbb{R}^{D\times d_h},\qquad
W^{K}_{\mathrm{sh}},W^{V}_{\mathrm{sh}}\in\mathbb{R}^{D\times d_h}.
\label{eq:mqsa_weights}
\vspace{-1ex}
\end{equation}

where, $W^{K}_{\mathrm{sh}}$ and $W^{V}_{\mathrm{sh}}$ are the shared projections. We then compute for $Q_h$, $K_{sh}$ and $V_{sh}$ as, $Q_h=XW_h^Q$ across all heads, while
$K_{\mathrm{sh}}=XW^{K}_{\mathrm{sh}},\ V_{\mathrm{sh}}=XW^{V}_{\mathrm{sh}}\in\mathbb{R}^{N\times d_h}$ is shared. The attention within each head is now $Y_h=\mathrm{Attn}(Q_h, K_{\mathrm{sh}}, V_{\mathrm{sh}})$ followed by $Y=\mathrm{Concat}_h(Y_h)W^O$.

From Equations~\ref{eq:mhsa_weights} and ~\ref{eq:mqsa_weights} we can conclude that by replacing the MHSA layers by MQA layer the number of parameters for the self-attention reduces by,
\vspace{-1ex}
\begin{equation}
\begin{aligned}
\underbrace{|W^K|+|W^V|}_{\text{MHSA}} &= 2D d_h H
\;\longrightarrow\;
\underbrace{|W^{K}_{\mathrm{sh}}|+|W^{V}_{\mathrm{sh}}|}_{\text{MQA}} = 2D d_h \\
&\Rightarrow\quad \mathbf{1/H}\ \text{reduction for } K,V .
\end{aligned}
\label{eq:mqsa_compression}
\end{equation}

We adopt this approach in ViTs and evaluate its effectiveness in a representative experiment shown in the lower right portion of  Figure~\ref{fig:2} using the training pipeline defined within Section~\ref{sec:kd}. Our results align with those of \cite{ainslie2023gqa}, demonstrating that given a pretrained checkpoint or a high-capacity teacher, MQA can approach the accuracy of its MHSA counterpart while reducing overall parameter count. 

\subsubsection{Ternary Quantization}
To aggressively compress the FFN layers found within \sys{}, we adapt the BitLinear layers from~\cite{ma2024era} which computes \(W_2A_8 \) output activations given 2-bit weights (\(W_2\)) and int8 activations (\(A_8\)). We modify this computation to operate over the full precision weight matrix \(W\) and patch embedding matrix \(A\) during training defined over the range \( W \in \mathbb{R}^{k \times n}, \; A \in \mathbb{R}^{m \times k} \). The quantized counterparts are then computed using absmean quantization\cite{ma2024era} for \(W_2\) and absmax quantization\cite{ma2024era} for \(A_8\) formally defined as
\vspace{-1ex}
\begin{equation}  
\begin{aligned}
W_2 = \mathrm{RoundClip}\Big( \frac{W}{\beta+\epsilon}, -1, 1 \Big), \quad
\beta = \frac{1}{kn} \textstyle \sum_{i=1,j=1}^{k,n} |W_{ij}|
\end{aligned}
\vspace{-2ex}
\end{equation}
and 
\vspace{-1ex}
\begin{equation}
\begin{aligned}
    A_8 = \mathrm{Clip}\Big( \frac{A}{\gamma+\epsilon}, -128, 127 \Big), \quad
\gamma = {\max(|x|)\over{127}}
\end{aligned}
\vspace{-1ex}
\end{equation}

Where $\epsilon$ represents a small floating point number. We modify $\gamma \in \mathbb{R}^{m}$ to be a vector of length determined by the the number of patch embeddings $m$, while $\beta \in \mathbb{R}$ remains as a single scalar representing the mean value over the entire weight matrix $W$. The scale factors \(\gamma,\beta\) are maintained within 16-bit precision during inference to ensure precise de-quantization for corresponding layers computed as $O=W_2A_8\times\gamma\times\beta$.

\begin{figure*}[!t]
    \centering
    \includegraphics[width=\textwidth]{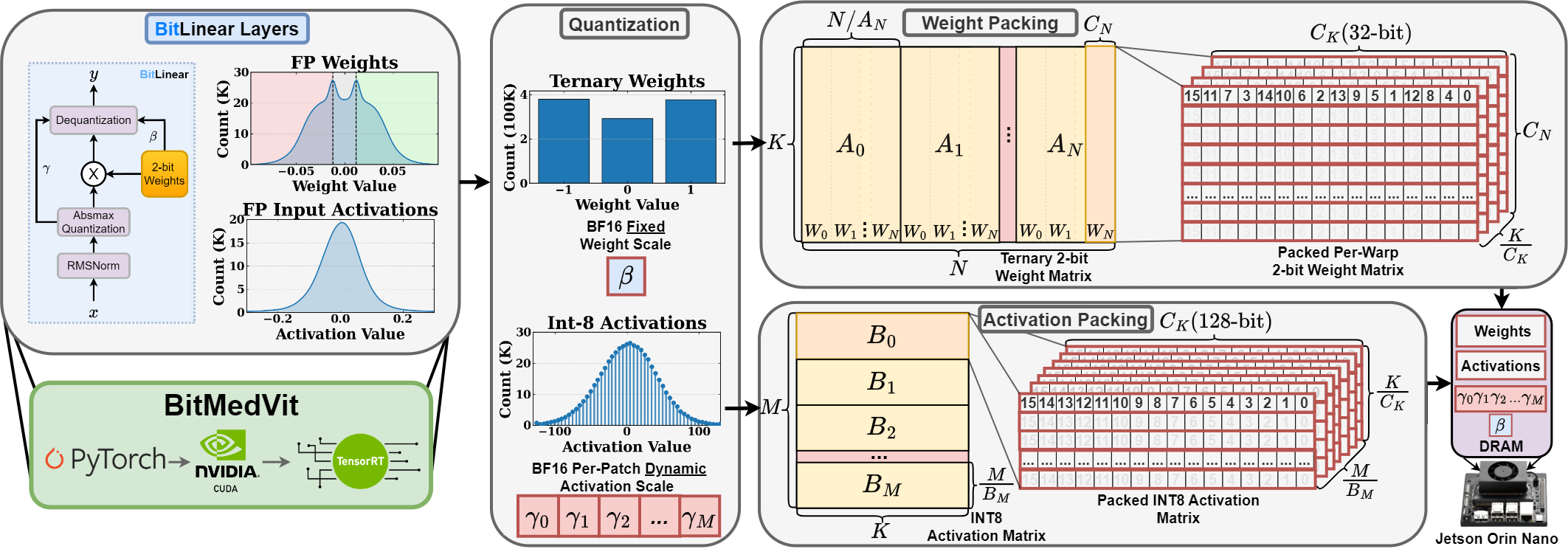}
    \vspace{-2ex}
    \caption{Bit Packing and custom hardware deployment strategy of \sys{} within the GPU of the Jetson Orin Nano. Full precision weights are statically quantized to 2-bit, packed into 32-bit column major words and activations dynamically quantized to int8 rowmajor 128-bit words converting the model from its Pytorch\cite{paszke2019pytorch} to the custom cuda implementation merged within TensorRT.}
    \label{fig:weight_packing}
    \vspace{-3ex}
\end{figure*}

\begin{algorithm}[]
\caption{CUDA Kernel for Blocked Matrix–Matrix Multiplication with 2-bit Weight Unpacking}
\label{algo:cuda_kernel}
\begin{algorithmic}[1]
\Require Weights $W$ (2-bit packed), patch input activations $P$ (int8), per-patch scale factors $\gamma$, weight scale factor $\beta$, block sizes $A$, $B$
\Ensure $O$ (dequantized, bfloat16) stored in global memory

\State \textbf{Init:} 
\Statex \hspace{\algorithmicindent} shared\_mem\_scale $\gets (\gamma \times \beta)$
\Statex \hspace{\algorithmicindent} $W_{\text{tile}} \gets$ weight start addr (index $A$)
\Statex \hspace{\algorithmicindent} $P_{\text{tile}} \gets$ activation start addr (index $B$)

\For{$k = 0$ \textbf{to} $K$ \textbf{step} tile\_size}
    \State \textbf{SyncLoad:} DRAM $\rightarrow$ registers $(W_{\text{packed}})$
    \State \textbf{AsyncLoad:} DRAM $\rightarrow$ shared\_mem $(P_{\text{tile}})$
    \State int8\_weights $\gets$ Unpack($W_{\text{packed}}$) $\rightarrow$ shared\_mem
    \State frag\_B $\gets$ \textbf{WMMA.Load}(int8\_weights)
    \State \textbf{SyncThreads}
    
    \For{$m = 0$ \textbf{to} $M/B - 1$}
        \State frag\_P $\gets$ \textbf{WMMA.Load}($P_{\text{tile}}[m]$)
        \State accum[m] $\gets$ \textbf{WMMA.MMA}(frag\_P, frag\_B)
    \EndFor
    
    \State \textbf{SyncThreads}
\EndFor

\State \textbf{Finalize:}
\Statex \hspace{\algorithmicindent} DRAM $\gets$ Dequantize(accum, shared\_mem\_scale)
\end{algorithmic}
\end{algorithm}

\subsection{Knowledge Distillation}\label{sec:kd}
Since the ternary quantized bitlinear layers require extensive training from scratch \cite{ma2024era,yuan2024vit} and MQA demonstrates effectiveness primarily when adapted from a pretrained model, we utilize MedViTv2~\cite{manzari2025medical} a state-of-the-art high-performing medical image classification ViT, achieving strong accuracy across medical imaging benchmarks. We train \sys{} by minimizing a composite loss function
\vspace{-2ex}

\begin{equation}
    \mathcal{L}_{\text{total}} = \lambda_{\text{cls}} \mathcal{L}_{\text{CE}} + \lambda_{\text{logits}} \mathcal{L}_{\text{KD}} + \lambda_{\text{feat}} \mathcal{L}_{\text{feat}}
    \vspace{-1ex}
\end{equation}

where \(\mathcal{L}_{\text{CE}}\) is the cross-entropy loss for classification, \(\mathcal{L}_{\text{KD}}\) denotes the Kullback--Leibler divergence aligning the student’s logits with the teacher’s, and \(\mathcal{L}_{\text{feat}}\) encourages alignment between intermediate feature representations. To facilitate this process, a trainable projection layer is incorporated during training to align the student’s feature dimensions with those of the teacher, in effect defining \sys{} to match the same patch size as MedViTv2.
\subsection{Model Optimization and Hardware Deployment}
\sys{} utilizes both ternary and full precision layers, necessitating a deployment strategy that optimizes for mixed precision within the GPU of the Orin Nano consisting of a custom CUDA kernel integrated within TensorRT\cite{nvidia_tensorrt}. To design our optimized kernel, The Jetson Orin Nano GPU is organized into a Graphics Processing Cluster (GPC), which contains four Texture Processing Clusters (TPCs), each with two Streaming Multiprocessors (SMs). Each SM includes four Tensor Cores capable of performing Warp Matrix Multiply (WMMA) operations at a minimum int8 precision, of varying dimensions. To maximize memory and compute bandwidth, operations are performed at the per-warp level, where a warp consists of 32 threads, with a maximum read transaction of int4 (128 bits) per thread. Each SM communicates with the 8GB external DRAM through a hierarchical memory system: a 192KB L1 cache per SM, a 4MB L2 cache shared among all SMs, and a 4MB system cache. Since DRAM reads introduce the highest latency, designing highly optimized kernels that perform compact, coalesced data accesses is essential to reduce L2 cache misses and maximize throughput.

\subsubsection{Weight Packing Strategy and WMMA Compatibility}
To optimize inference of \sys{}, we redesign the weight packing scheme from the original BitNet~\cite{bitnet_gpu} kernel to align with  operations for maximizing onboard Tensor Core utilization. Weights are arranged into matrices of size \(8 \times 32 \times 16\) (M \(\times\) N \(\times\) K). Within this layout, weights are packed into column major \(32 \times 16\) fragments, where each 32-bit memory word encodes sixteen ($C_K$) 2-bit weight values with activations quantized to 8-bit integers grouped into 128-bit (int4) fragments, allowing sixteen activation values to be loaded per memory transaction. as shown in Figure~\ref{fig:weight_packing}. This compact packing reduces memory traffic and allows efficient unpacking via optimized bit masking and shifting during inference. In addition the 32 outputs of $N$ per input patch $M$ enables coalesced and consecutive 16-bit (BF16) output write-backs during de-quantization.

\subsubsection{CUDA Kernel Integration}
As outlined within Figure~\ref{fig:weight_packing} Our custom CUDA kernel parallelizes computation using a two-dimensional grid of size \(A \times B\), where \(A\) corresponds to output channels and \(B\) to input patches divided among each of the onboard SMs. Each thread block contains multiple 32-thread warps, with threads accessing distinct output elements acting on the same activation fragment. To minimize runtime overhead, decoded weights are unpacked once per inference and stored in shared memory within each thread block. Activations, which have lower temporal reuse, are asynchronously loaded into shared memory, skipping past the L1 cache maintaining a continuous data streaming pipeline and enabling weight decoding during activation loading. Algorithm~\ref{algo:cuda_kernel} outlines this functionality and the overall per-block kernel execution.

\begin{figure}[h]
    \vspace{-2ex}
    \centering
    \includegraphics[width=1.\linewidth]{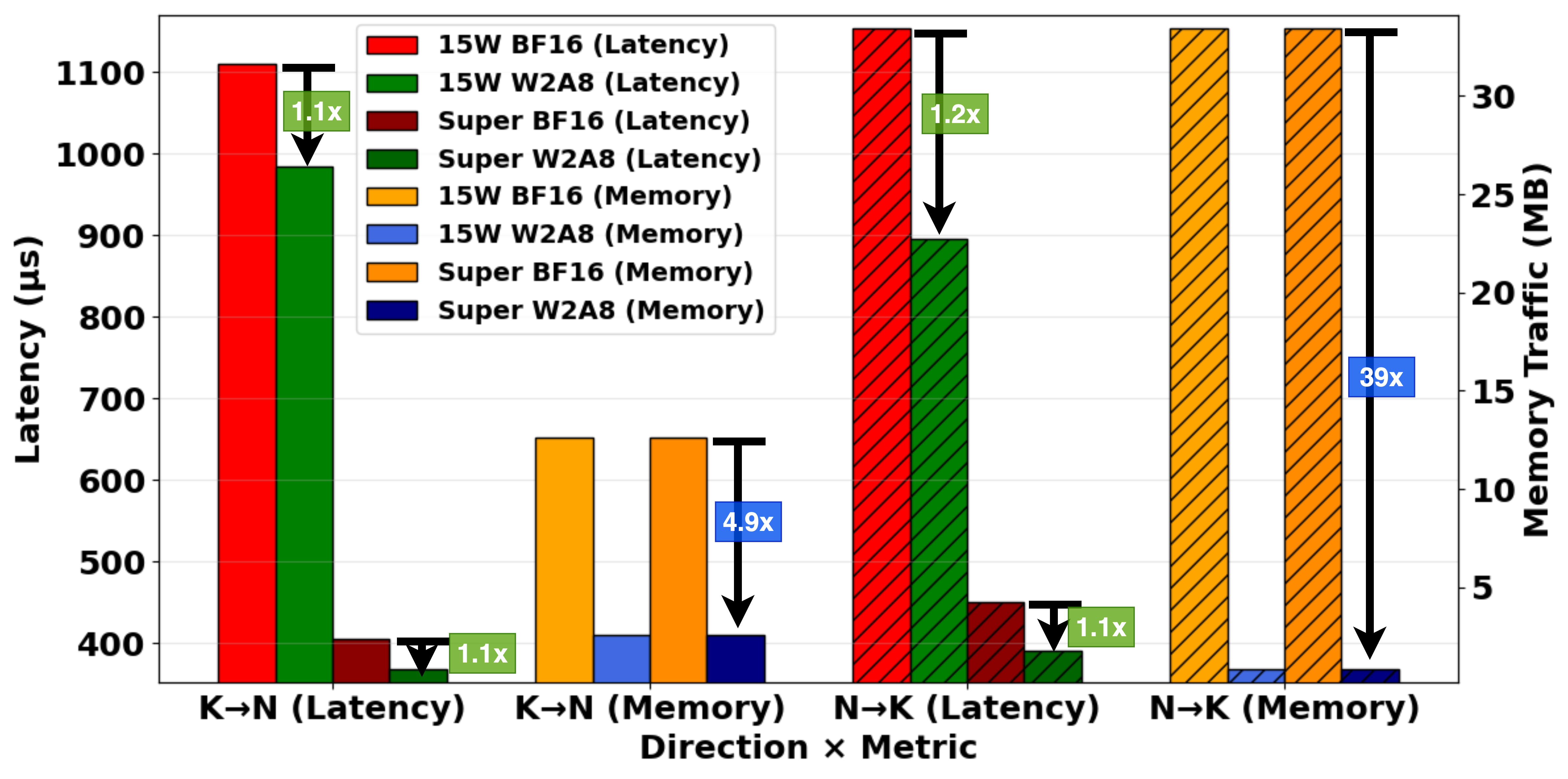}
    \vspace{-5ex}
    \caption{Latency and memory read traffic comparing the PyTorch\cite{paszke2019pytorch} BF16 kernel to our optimized W2A8 implementation for two varying workloads. Results on the left show the latency and memory read transactions for the K-N layer and results on the right show the N-K layer within the \sys{} FF network. Two power modes, 15W and Super (25W) mode are used to determine performance under varying clock speeds.}
    \label{fig:kernel_prof}
    \vspace{-2ex}
\end{figure}

Using the Nvidia Nsight Compute \cite{nvidia_nsight_compute} and Nsight Systems \cite{nvidia_nsight_systems} profiling tools, we extensively evaluate the efficiency of the custom kernel implementation as well as overall inference time for the full model architecture. Profiling the linear layers in \sys{} as shown in Figure ~\ref{fig:kernel_prof} shows a 4.9 $\times$ to 39$\times$ reduction in weight traffic (bytes moved). This shifts most accesses to on-chip memory and minimizes DRAM transactions, aligning with the power/latency gains.


\subsubsection{TensorRT Integration for Efficient Deployment}
We deploy \sys{} as an end-to-end solution by integrating our custom CUDA kernel into NVIDIA TensorRT. Since TensorRT doesn’t natively support quantized/compressed BitLinear layers, the plugin registers the layer, declares I/O shapes, and specifies supported precisions enabling building of engines with the new operation. On the Orin Nano, TensorRT had limited support for dynamic INT8 activations and would upcast them to FP16 at the plugin boundary. To accommodate this, we added kernel variants that accept BF16 activations and scales, and due to the limited BF16 support on the Orin Nano, FP16 activations and scales. 

\begin{table*}
\caption{Performance of \sys{} across MedMNIST2D datasets in comparison with state-of-the-art classification models. The teacher model MedViTv2-L and \sys{} are highlighted. Validation accuracy is reported for fair comparison, while test accuracy for each dataset is indicated with a $*$.}
\vspace{-2ex}
\centering
\begin{tabular}{c|cc|cc|cc|cc|cc|cc}
\toprule
 & \multicolumn{2}{c|}{PathMNIST} & \multicolumn{2}{c|}{ChestMNIST} & \multicolumn{2}{c|}{DermaMNIST} & \multicolumn{2}{c|}{OCTMNIST} & \multicolumn{2}{c|}{PneumoniaMNIST} & \multicolumn{2}{c}{RetinaMNIST} \\ 
\multirow{-2}{*}{Model} & ACC & AUC & ACC & AUC & ACC & AUC & ACC & AUC & ACC & AUC & ACC & AUC \\ \midrule
ResNet50 (224)~\cite{he2016deep} & 89.2 & 98.9 & 94.8 & 77.3 & 73.1 & 91.2 & 77.6 & 95.8 & 88.4 & 96.2 & 51.1 & 71.6 \\ \midrule
MedMamba-S~\cite{yue2024medmamba} & 95.5 & 99.7 & - & - & 75.8 & 92.4 & 92.9 & 99.1 & 93.6 & 97.6 & 54.5 & 71.8 \\
MedMamba-B~\cite{yue2024medmamba} & 96.4 & 99.9 & - & - & 75.7 & 92.5 & 92.7 & 99.6 & 92.5 & 97.3 & 55.3 & 71.5 \\ \midrule
MedViTv2-S~\cite{manzari2025medical} & 96.5 & 99.8 & 96.4 & 80.3 & 79.2 & 94.6 & 94.2 & 99.4 & 96.5 & 99.6 & 56.2 & 78.0 \\
MedViTv2-B~\cite{manzari2025medical} & 97.0 & 99.9 & 96.4 & 81.5 & 80.8 & 94.8 & 94.4 & 99.6 & 96.9 & 99.6 & 57.5 & 78.3 \\
\rowcolor[HTML]{ECEAEA} 
MedViTv2-L~\cite{manzari2025medical} & \begin{tabular}[c]{@{}c@{}}97.7\\  *93.0\end{tabular} & \begin{tabular}[c]{@{}c@{}}99.9\\  *100.0\end{tabular} & \begin{tabular}[c]{@{}c@{}}96.7\\  *93.6\end{tabular} & \begin{tabular}[c]{@{}c@{}}82.3\\  *75.6\end{tabular} & \begin{tabular}[c]{@{}c@{}}81.7\\  *83.0\end{tabular} & \begin{tabular}[c]{@{}c@{}}95.0\\  *93.0\end{tabular} & \begin{tabular}[c]{@{}c@{}}95.2\\  *94.8\end{tabular} & \begin{tabular}[c]{@{}c@{}}99.6\\  *100.0\end{tabular} & \begin{tabular}[c]{@{}c@{}}97.3\\  *97.0\end{tabular} & \begin{tabular}[c]{@{}c@{}}99.7\\  *99.0\end{tabular} & \begin{tabular}[c]{@{}c@{}}57.8\\  *50.3\end{tabular} & \begin{tabular}[c]{@{}c@{}}78.5\\  *75.8\end{tabular} \\ \toprule
\rowcolor[HTML]{ECEAEA} 
\textbf{BitMedViT (ours)} & \textbf{\begin{tabular}[c]{@{}c@{}}99.0\\  *91.1\end{tabular}} & \textbf{\begin{tabular}[c]{@{}c@{}}100.0\\  *99.2\end{tabular}} & \textbf{\begin{tabular}[c]{@{}c@{}}94.0\\  *93.7\end{tabular}} & \textbf{\begin{tabular}[c]{@{}c@{}}72.0\\  *71.5\end{tabular}} & \textbf{\begin{tabular}[c]{@{}c@{}}79.0\\  *76.2\end{tabular}} & \textbf{\begin{tabular}[c]{@{}c@{}}95.0\\  *93.6\end{tabular}} & \textbf{\begin{tabular}[c]{@{}c@{}}95.0\\  *85.7\end{tabular}} & \textbf{\begin{tabular}[c]{@{}c@{}}99.0\\  *98.8\end{tabular}} & \textbf{\begin{tabular}[c]{@{}c@{}}86.0\\  *88.0\end{tabular}} & \textbf{\begin{tabular}[c]{@{}c@{}}98.0\\  *95.0\end{tabular}} & \textbf{\begin{tabular}[c]{@{}c@{}}53.0\\  *51.3\end{tabular}} & \textbf{\begin{tabular}[c]{@{}c@{}}81.0\\  *73.3\end{tabular}} \\ \toprule

\toprule
 & \multicolumn{2}{c|}{BreastMNIST} & \multicolumn{2}{c|}{BloodMNIST} & \multicolumn{2}{c|}{TissueMNIST} & \multicolumn{2}{c|}{OrganAMNIST} & \multicolumn{2}{c|}{OrganCMNIST} & \multicolumn{2}{c}{OrganSMNIST} \\ 
\multirow{-2}{*}{Model} & ACC & AUC & ACC & AUC & ACC & AUC & ACC & AUC & ACC & AUC & ACC & AUC \\ \midrule
ResNet50 (224)~\cite{he2016deep} & 84.2 & 86.6 & 95.0 & 99.7 & 68.0 & 93.2 & 94.7 & 99.8 & 91.1 & 99.3 & 78.5 & 97.5 \\ \midrule
MedMamba-S~\cite{yue2024medmamba} & 85.3 & 80.6 & 98.4 & 99.9 & - & - & 95.9 & 99.9 & 94.4 & 99.7 & 83.3 & 98.4 \\
MedMamba-B~\cite{yue2024medmamba} & 89.1 & 84.9 & 98.3 & 99.9 & - & - & 96.4 & 99.9 & 94.3 & 99.8 & 83.4 & 98.3 \\ \midrule
MedViTv2-S~\cite{manzari2025medical} & 89.5 & 94.7 & 98.5 & 99.9 & 70.5 & 93.9 & 96.6 & 99.9 & 95.0 & 99.8 & 83.9 & 98.6 \\
MedViTv2-B~\cite{manzari2025medical} & 90.4 & 94.9 & 98.5 & 99.9 & 71.1 & 94.2 & 96.9 & 99.9 & 95.3 & 99.8 & 84.4 & 98.7 \\
\rowcolor[HTML]{ECEAEA} 
MedViTv2-L~\cite{manzari2025medical} & \begin{tabular}[c]{@{}c@{}}91.0\\  *86.5\end{tabular} & \begin{tabular}[c]{@{}c@{}}95.3\\  *92.5\end{tabular} & \begin{tabular}[c]{@{}c@{}}98.7\\  *98.5\end{tabular} & \begin{tabular}[c]{@{}c@{}}99.9\\  *100.0\end{tabular} & \begin{tabular}[c]{@{}c@{}}71.6\\  *75.7\end{tabular} & \begin{tabular}[c]{@{}c@{}}94.3\\  *95.7\end{tabular} & \begin{tabular}[c]{@{}c@{}}97.3\\  *84.6\end{tabular} & \begin{tabular}[c]{@{}c@{}}99.9\\  *98.8\end{tabular} & \begin{tabular}[c]{@{}c@{}}96.1\\  *86.7\end{tabular} & \begin{tabular}[c]{@{}c@{}}99.9\\  *98.9\end{tabular} & \begin{tabular}[c]{@{}c@{}}85.1\\  *82.7\end{tabular} & \begin{tabular}[c]{@{}c@{}}98.7\\  *98.3\end{tabular} \\ \toprule
\rowcolor[HTML]{ECEAEA} 
\textbf{BitMedViT (ours)} & \textbf{\begin{tabular}[c]{@{}c@{}}87.0\\  *82.1\end{tabular}} & \textbf{\begin{tabular}[c]{@{}c@{}}91.0\\  *82.6\end{tabular}} & \textbf{\begin{tabular}[c]{@{}c@{}}97.0\\  *97.5\end{tabular}} & \textbf{\begin{tabular}[c]{@{}c@{}}100.0\\  *99.9\end{tabular}} & \textbf{\begin{tabular}[c]{@{}c@{}}64.0\\  *63.8\end{tabular}} & \textbf{\begin{tabular}[c]{@{}c@{}}92.0\\  *91.9\end{tabular}} & \textbf{\begin{tabular}[c]{@{}c@{}}98.0\\  *90.2\end{tabular}} & \textbf{\begin{tabular}[c]{@{}c@{}}100.0\\  *99.5\end{tabular}} & \textbf{\begin{tabular}[c]{@{}c@{}}96.0\\  *88.4\end{tabular}} & \textbf{\begin{tabular}[c]{@{}c@{}}100.0\\  *99.2\end{tabular}} & \textbf{\begin{tabular}[c]{@{}c@{}}81.0\\  *74.1\end{tabular}} & \textbf{\begin{tabular}[c]{@{}c@{}}99.0\\  *97.4\end{tabular}}\\\toprule

\end{tabular}
\label{tab:accuracy comparisons}
\vspace{-2ex}
\end{table*}
\begin{figure}[t]
    \centering
    \includegraphics[width=\linewidth]{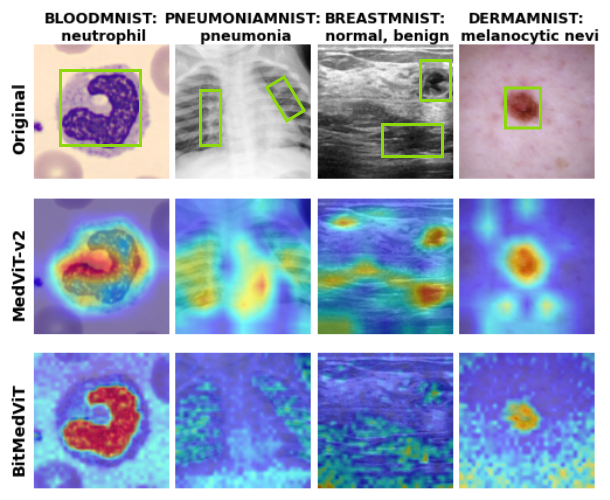}
    \vspace{-5ex}
    \caption{Visualized GradCam\cite{selvaraju2017grad} for the MedViT-v2-L teacher and \sys{} across four representative datasets found within MedMNIST. Green boxes represent the manually placed relevant regions for diagnosis.}
    \label{fig:medmnist_results}
    \vspace{-3ex}
\end{figure}

\section{Results}
\vspace{-1ex}
\subsection{\sys{} Accuracy Benchmarking}
MedMNIST2D~\cite{medmnistv1, medmnistv2} serves as the dataset used to evaluate the performance of \sys{} for medical image classification   spanning 12 2D datasets of varying modalities and complexity. Using the pretrained MedViTv2-L teacher, we apply the distillation strategy outlined in Section~\ref{sec:kd} across each dataset. Figure~\ref{fig:medmnist_results} presents attention maps generated using GradCam~\cite{selvaraju2017grad} for four representative images, which, while exhibiting increased noise, demonstrate stronger focus on diseased regions. These results highlight the ability of \sys{} to attend to clinically relevant regions despite aggresive reduction in parameters and weight expressivity.

Furthermore, we compare overall accuracy and ROC AUC against state-of-the-art medical image classification models, as shown in Table~\ref{tab:accuracy comparisons}. Across the MedMNIST benchmark, \sys{} achieves competitive accuracy relative to leading models, attaining perfect or near-perfect AUC scores on multiple datasets, including PathMNIST, BloodMNIST, and OrganCMNIST. Notably, even after applying aggressive compression strategies, \sys{} maintains performance close to the teacher, with only a 3\% decrease in average validation accuracy (86\% vs. 89\%) and a similar reduction on the test set (82\% vs.\ 85\%), demonstrating effective model compression while preserving competitive performance.

\subsection{Hardware Deployment}

\begin{figure}[]
    \centering
    \vspace{-1ex}
    \includegraphics[width=\linewidth]{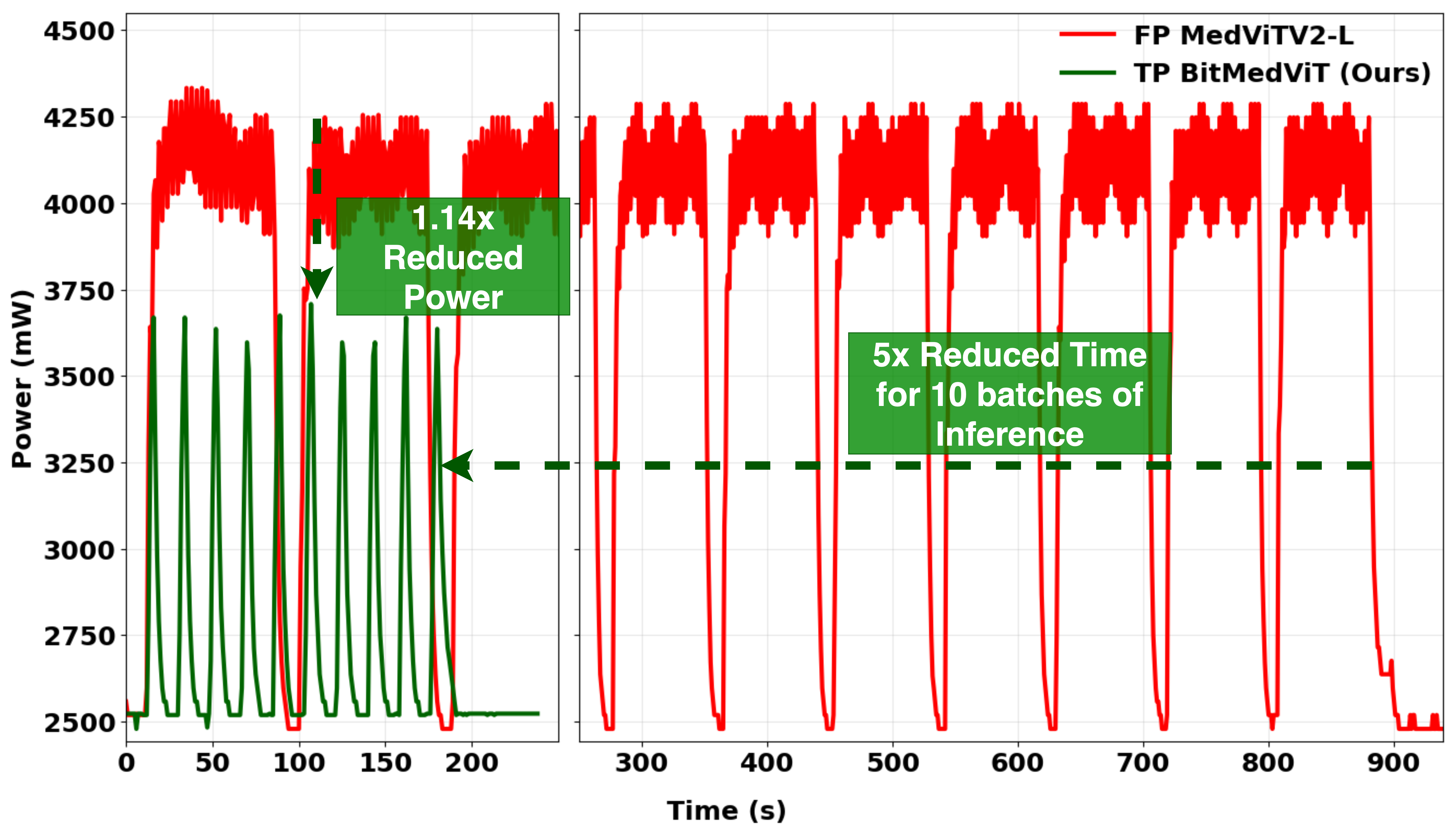}
    \caption{Jetson Orin Nano Power Versus Time for the Ternary Precision (TP) BitMedVit(Ours) and Full precision (FP) MedVitV2-L teacher. \sys{} achieves nearly a $1.14\times$ reduction in power while achieving nearly a $5\times$ reduction in latency for processing 10 batches of 50 inferences. }
    \label{fig:power_graph}
    \vspace{-4.6ex}
\end{figure}

\begin{table*}
\centering
\caption{Hardware Comparison Against state of the art medical image classification models deployed within the Jetson Orin Nano. Values with * are recomputed with our metrics}
\vspace{-2ex}
\begin{tabular}{c|c|c|c|c|c|c|c|c|c}
\toprule
Work & \begin{tabular}[c]{@{}c@{}}Model \\ Type\end{tabular} & Precision & \begin{tabular}[c]{@{}c@{}}Parameters \\ (M)\end{tabular} & \begin{tabular}[c]{@{}c@{}}Model \\ Size \\ (MB)\end{tabular} & \begin{tabular}[c]{@{}c@{}}Operations \\ (GOPs)\end{tabular} & \begin{tabular}[c]{@{}c@{}}Power \\ (W)\end{tabular} & \begin{tabular}[c]{@{}c@{}}Latency\\ (ms)\end{tabular} & \begin{tabular}[c]{@{}c@{}}Performance \\ (GOPs / sec)\end{tabular} & \begin{tabular}[c]{@{}c@{}}Energy \\ Efficiency \\ (GOPs / J)\end{tabular} \\ \toprule
\begin{tabular}[c]{@{}c@{}}MedViTv2-L (Baseline)~\cite{manzari2025medical}\end{tabular} & ViT & Float32 & 117.26 & 447.71 & 12.9 & 4.25 & 366.63 & 35.3 & 8.31 \\ \toprule
MedMambaLite-ST~\cite{aalishah2025medmambalite} & \begin{tabular}[c]{@{}c@{}}Mamba\end{tabular} & Float32 & 0.63 & 2.4 & 0.15 & 2.7 & 13.03 & 11.84 & *4.39 \\ \toprule
\rowcolor[HTML]{ECEAEA} 
\textbf{\begin{tabular}[c]{@{}c@{}}BiTMedViT (ours)\end{tabular}} & \textbf{ViT} & \textbf{W2A8} & \textbf{8.65} & \textbf{10.5} & \textbf{11.53} & \textbf{3.72} & \textbf{16.88} & \textbf{683.06} & \textbf{183.62} \\ \toprule
\end{tabular}
\label{tab:hw_comp}
\vspace{-4ex}
\end{table*}

We deploy \sys{} within the Orin Nano and compare against (i) full-precision MedViTv2-L\cite{manzari2025medical} baseline and (ii) MedMambaLite\cite{aalishah2025medmambalite}. with hardare results summarized in Table~\ref{tab:hw_comp}. \sys{} achieves 16.88\,ms latency per inference versus 366.63\,ms for the MedViTv2-L baseline ($\approx$21.7$\times$ faster). This corresponds to a 683.06\,GOPs/sec throughput, 19.4$\times$ higher than the baseline. When compared to MedMambaLite\cite{aalishah2025medmambalite}, our implementation is $57.7\times$ better in terms of throughput while being $\approx42\times$ more energy efficient. Figure~\ref{fig:power_graph} shows the power trace for executing an identical workload on the teacher and \sys{} on the Orin Nano platform: 50 inferences followed by 3 seconds of idle time, repeated 10 times. Compared to the teacher, \sys{} achieves a $5\times$ reduction in total inference time and a $1.14\times$ lower peak power consumption.
\section{Conclusion}\label{sec:conclusion}
In this work, we presented \sys{}, a ternary-quantized ViT for efficient, real-time medical image classification on the edge. By integrating Multi-Query Attention, knowledge distillation, and hardware-aware CUDA \& TensorRT optimization, BitMedViT achieves 86\% accuracy on MedMNIST only 3\% below its teacher while reducing parameters by $13.6\times$, model size by $43\times$, and memory transfers by $39\times$, at a 92\% L2 cache hit rate and performing at 183.62 GOPs/J, 22$\times$ that of MedViTv2-L~\cite{manzari2025medical} and $42\times$ that of MedMambaLite-ST~\cite{aalishah2025medmambalite}. These results demonstrate that extreme-precision quantization, combined with architectural and deployment co-design, enables SOTA ViT performance within the strict compute and memory constraints of clinical edge devices.

Future work will explore mixed-precision training techniques to dynamically adjust bit precisions between layers, alongside adaptive quantization based on input modality or task complexity. In addition, expanding \sys{} to CPU-only and specialized low-power devices will further enhance its deployability and accessibility across diverse clinical environments and devices
\vspace{-2ex}

\bibliographystyle{plainnat}
\begin{small}
\bibliography{references}
\end{small}

\end{document}